\documentclass[structabstract]{aa}  
\usepackage{graphicx}
\usepackage{txfonts}
\begin{document}
   \title{Constraints on the exosphere of CoRoT-7b
\thanks{based on observations obtained at the European Southern
    Observatory at Paranal, Chile in program 384.C-0820(A)}}
\authorrunning{Guenther et al.}
\titlerunning{The exosphere of CoRoT-7b}
\author{E.W. Guenther\inst{1,2}, J. Cabrera\inst{3,4}, A. Erikson
  \inst{3}, M. Fridlund \inst{5}, H. Lammer \inst{6}, A. Mura \inst{7}, 
  H. Rauer \inst{3,9}, J. Schneider
  \inst{4}, M. Tulej\inst{8}, Ph. von Paris, \inst{3}, P. Wurz\inst{8}
          }
   \institute{Th\"uringer Landessternwarte Tautenburg,
              Sternwarte 5, D-07778 Tautenburg, Germany\\
              \email{guenther@tls-tautenburg.de}
         \and
              Instituto de Astrof\'\i sica de Canarias,
              C/V\'\i a L\'actea, s/n,
              E38205 -- La Laguna (Tenerife), Spain
         \and
              Institute of Planetary Research, German Aerospace Center, 
              Rutherfordstra\ss e 2, 12489 Berlin, Germany
         \and
              LUTH, Observatoire de Paris, CNRS, Universit\'e Paris Diderot, 
              5 place Jules Janssen, 92190 Meudon, France
        \and
               Research and Scientific Support Department, 
              ESTEC/ESA, PO Box 299, 2200 AG Noordwijk, The Netherlands
        \and
              Space Research Institute, Austrian Academy of Science, 
              Schmiedlstra\ss e. 6, 8042 Graz, Austria 
        \and
              Istituto di Fisica dello Spazio Interplanetario-CNR, Rome, Italy
        \and
              Physics Institute, University of Bern, Bern, Switzerland
        \and
              Zentrum f\"ur Astronomie
              und Astrophysik (ZAA) Technische Universit\"at Berlin (TUB)
              Hardenbergstra\ss e 36 10623 Berlin, Germany
            }

   \date{Received April 26, 2010; accepted September 23, 2010}

  \abstract 
% context heading (optional) % {heading} leave it empty if necessary 
{The small radius and high density of CoRoT-7b implies that this
  transiting planet belongs to a different species than all transiting
  planets previously found. Current models suggest that
  this is the first transiting rocky planet found outside the solar
  system.  Given that the planet orbits a solar-like star at a
  distance of only 4.5 $R_{*}$, it is expected that material 
  released from its surface  may then form an exosphere.}
{We constrain the properties of the exosphere by observing
  the planet in- and out-of-transit.  Detecting of the exosphere of
  CoRoT-7b would for the first time allow to study the material
  originating in the surface of a rocky extrasolar planet.  We
  scan the entire optical spectrum for any lines originating from the
  planet, focusing particularly on spectral lines such as those 
  detected in Mercury, and Io in our solar system.}
{Since lines originating in the exosphere are expected to
  be narrow, we observed CoRoT-7b at
  high resolution with UVES on the VLT.  By subtracting the two
  spectra from each other, we search for emission and absorption lines
  originating in the exosphere of CoRoT-7b.}
{In the first step, we focus on Ca\,I, Ca\,II, Na, because these lines
  have been detected in Mercury.  Since the signal-to-noise ratio
  (S/N) of the spectra is as high as 300, we derive firm upper limits
  for the flux-range between $1.6\times 10^{-18}$ and $3.2\times 10^{-18}$
  $W\,m^{-2}$. For CaO, we find an upper limit of $10^{-17}$
  $W\,m^{-2}$. We also search for emission lines originating in the
  plasma torus fed by volcanic activity and derive upper limits for
  these lines. In the whole spectrum we finally try to identify
  other lines originating in the planet.}
{Except for CaO, the upper limits derived correspond to $2-6\times
  10^{-6}$ $L_{*}$, demonstrating the capability of UVES to detect
  very weak lines. Our observations certainly exclude the extreme
  interpretation of data for CoRoT-7b, such as an exosphere that emits
  2000 times as brightly as Mercury. }
   \maketitle
%
%________________________________________________________________

\section{Introduction}

Studies of transiting extra-solar planets are of key importance for
understanding the nature of planets, because it is possible to derive
their mass, diameter, and hence their density.  Observations of
transits even allow us to detect the atmosphere of the planets.  Up to
now, studies of the atmospheres of extrasolar planets have
concentrated on giant gaseous planets.  The detection of transiting
planets of low mass and small radius now permits us
to examine the atmosphere of planets that are possibly rocky
planets.
 
\object{CoRoT-7b} is an extrasolar planet of a radius of
$1.58\pm0.10$ $R_{Earth}$ orbiting a G9V star at a distance of only
4.5 $R_*$ (original values: L\'eger et al. \cite{leger09}; new values:
Bruntt et al. \cite{bruntt10}). The mass is $6.9\pm1.4\,M_{Earth}$,
and the density is $\rho =9.6 \pm 2.7\,g\,cm^{-3}$ (original value:
Queloz et al. \cite{queloz09}; new value: Hatzes et
al. \cite{hatzes10}). The density is thus even higher than Earth
($\rho =5.515\,g\,cm^{-3}$), Venus ($\rho =5.243\,g\,cm^{-3}$), or
Mercury ($\rho =5.427\,g\,cm^{-3}$).

Although the density of this planet even exceeds those of the rocky
planets of our solar system, it may not necessarily be a rocky
planet. However, its low mass, small radius, and high density indicate
that it certainly belongs to a different species than the gaseous
planets studied so far. It would thus be exiting to determine its
surface composition, and assess whether it has an atmosphere or not.

Future instruments such as SEE-COAST (now called SPICES) are designed
to study the atmospheres and surface compositions of planets down to a
few Earth-radii (Schneider et al. \cite{schneider09}).  The
instrumentation study SEARCH (Mohler et al. \cite{mohler10})
demonstrates that an 8-m-class space telescope equipped with a highly
sophisticated spectropolarization device would be able to study the
atmospheres and surface composition of extrasolar planets of similar
size to the Earth. While the prospects for such studies are high, it
will take decades until such missions are operate. However, there is
already a fair chance of detecting material released from the surface
of rocky extrasolar planets using existing instrumentation.

If a sufficient quantity of material is released from the surface of
the planet into space, its surface composition may be studied by
performing spectroscopy when it is both in- and out-of-transit.

While the exosphere of gaseous extrasolar planets has been
detected in hydrogen as well as heavier elements (Vidal-Madjar et
al. \cite{vidal03}; Vidal-Madjar et al. \cite{vidal04}; Lecavelier des
Etangs et al. \cite{lecavalier10}; Linsky et al. \cite{linsky10};
Holmstr{\"o}m, et al. \cite{holmstrom08}; Eckenb\"ack et al.
\cite{eckenbaeck10}; Fossati \cite{fossati10}), no attempts have yet
been made to detect the exosphere of a rocky extrasolar planet.  

Our observations represent the first attempt to detect the exosphere
of a rocky extrasolar planet. We present UVES observations taken in-
and out-of-transit that we use to search for emission lines
originating in material released from the surface of the planet.

\section{What we might expect to find}

Whether or not material from the surface of the planet is released
depends not only on the properties of the planet but also those of the
star. The host star has a an effective temperature $T_{eff}=5250\pm
60\,K$, a mass of $M_{*}=0.91\pm0.03\,M_\odot$, and a radius of
$R_{*}=0.82\pm0.04\,R_\odot$ (Bruntt et al. \cite{bruntt10}).
Although the rotation period of the host star is at 23 days quite
close to that of the Sun, the star is more active than the Sun. The
total chromospheric radiative loss in the Ca\,II\,H,K lines in units
of the bolometric luminosity is $\log R _{\rm HK} = - 4.601 \pm 0.05$,
which implies a significantly higher activity level than the Sun, for
which $\log R _{\rm HK} = - 4.901$ (Baliunas, Sokoloff \& Soon
\cite{baliunas96}). On the basis of this activity index, the the age
of the star is estimated to be in the range 1.2 - 2.3 Gyr.  Given that
the star is quite similar to the Sun, apart from being slightly more
active, it is reasonable to assume that its stellar wind is also
quite similar to that of the Sun. Because the star is an active
G-star, we can safely assume that there is a stellar wind that will
interact with the planets.

Before we go ahead and try to detect the material released from the
surface, we discuss how likely the planet is to be rocky.  Until now,
we know about the planets mass and radius, not its composition.
Assuming a rocky planet, Valencia et al. (\cite{valencia10}) could
reproduce the original values of the radius and mass of $1.68\pm0.10$
$R_{Earth}$ and $4.8\pm0.8$ $M_{Earth}$, respectively, given in
L\'eger et al. (\cite{leger09}) and Queloz et al. (\cite{queloz09}).
However, to reproduce the relatively low density given in these
articles they had to assume that the planet is significantly depleted
in iron compared to the Earth. The authors argue that this is unlikely
and that it is more likely that it has the same composition as the
Earth. If this were the case, the mass and radius would be $5.6$
$M_{Earth}$ and $1.59$ $R_{Earth}$, respectively.  Interestingly, the
new values of mass and radius are $6.9\pm1.4$ $M_{Earth}$ and
$1.58\pm0.10$ $R_{Earth}$, respectively, which are perfectly
consistent with those of a rocky planet with the composition of the
Earth.  Using realistic values for the heating efficiencies, and also
taking the evolution of the EUV flux of the host star into account,
Leitzinger et al. (\cite{leitzinger10}) showed that CoRoT-7b was
always a rocky planet and is not the eroded core of a gas giant. Given
all these results, it is justified to assume that it \object{CoRoT-7b}
is a rocky planet.

Given that we do not have a planet like \object{CoRoT-7b} in our solar
system, the structure, surface composition, and how much material is
released from its surface is not known. For this reason, we scan the
whole optical spectrum for lines originating in the
planet. Nevertheless, it makes sense to discuss which lines we might
expect. \object{CoRoT-7b} is certainly not like either Mercury or Io
but these objects do have some similarities to \object{CoRoT-7b}.  We
use these analogies to discuss which lines we should be looking for.

In Sect. 2 we discuss the species we expect to find.  In Sect. 3, we
report on the observations, in Sect. 4 we present the limits derived
for Na, Ca, $\rm Ca^{+}$, CaO, $\rm S^{+}$, $\rm S^{2+}$, and $\rm
O^{2+}$, and in Sect. 4 we discuss the results obtained.

\subsection{The Mercury analogy}

To some extent, the local environment of \object{CoRoT-7b} is an
extreme version of Mercury, as its distance to the host star is 23
times smaller than that of Mercury.  Hence the planet receives about
250-370 times more radiation than Mercury. In addition, the stellar wind is
more intensive than at Mercury, because the host star is more active
than the Sun, and the planet closer to its host star.  Could it
also have a Mercury-like exosphere, and if so what would we expect to
observe?

The gaseous envelope of Mercury was discovered by the Mariner 10
spacecraft.  Since Mercury's envelope is collision less, it resembles
an exosphere where the exobase is coincident with the planet's
surface.  An exosphere is the collision less, outermost layer of an
atmosphere of the planet where atoms, ions, or molecules can escape
into space.  Mariner 10 detected UV emission from the exosphere of
three atomic elements: H, He, and O (Broadfoot et
al. \cite{broadfoot76}). Three other elements (Na, K, and Ca) were
later discovered by ground-based observations (Potter \& Morgan
\cite{potter85} and Bida, Killen \& Morgan \cite{bida00}). The
exosphere of Mercury is extremely thin. The surface pressure is only
$10^{-15}\,bar$, and the total mass $\leq 1000\,kg$, which is tiny
compared to the atmosphere of the Earth with its mass of
$5\times10^{18}\,kg$.  It is believed that the exosphere of Mercury is
derived in large part from the surface materials (Wurz \& Lammer,
\cite{wurz03}; Wurz et al. \cite{wurz09}).

A combination of for instance impact vaporization (from in-falling
material), volatile evaporation, photon-stimulated desorption, and
sputtering, releases material from the surface to form the
exosphere. The ground-based observations by Bida, Killen \& Morgan
(\cite{bida00}) reveal Ca\,I emission in the resonance line at 4226.74
\AA .  Observations performed during the transit of Mercury in 2003
detected additionally sodium emission lines (Na I $D_{1,2}$), which
allowed Schleicher et al. (\cite{schleicher04}) to trace the extent of
the exosphere of Mercury above the planet's limb.  Potter et
al. (\cite{potter02}) observed a sodium-tail in the anti-sunward
direction extending ten's of planetary radii.

Mura et al. (\cite{mura10}) applied their model of
Mercury to \object{CoRoT-7b} and found that surface elements should
in the case be released into space.  By analogy with Mercury, the authors
postulate that there is a sodium-tail extending ten's of planetary
radii into the anti-stellar direction. According to these authors, the
radiation pressure on the sodium atoms is about a factor of 100 higher
than the gravitational force, which causes the sodium-tail to be
almost perfectly aligned with the star-planet axis and the 
cross-section of the sodium-tail to be only slightly larger than the planet
itself. In addition, Mura et al. (\cite{mura10}) postulate the
existence of a $Ca^+$-ion tail. Exospheric Ca is ionized very rapidly
by the stellar photon field and the $Ca^+$-ions are guided by the
stellar wind. Given the high transverse velocity of the planet of 218
$km\,s^{-1}$ with respect to the stellar wind (estimated to be around
200 $km\,s^{-1}$), which flows radially away from the star, the ion
tail is expected to be inclined by about $45^o$ with respect to the
star-planet line. Thus, during transit the Ca+ tail is also inclined
by $45^o$ to the line of sight. According to these authors, the
Ca\,II lines are seen in absorption during the transit.  They find
that the coma has an extension of several $R_\odot$. We conclude
that \object{CoRoT-7b} may have an exosphere and given the extreme
conditions on \object{CoRoT-7b} the exosphere may also be
detectable.

We note that the a cometary tail of a hot Jupiter \object{HD209458\,b}
has apparently already been detected (Linsky et al. \cite{linsky10}).

\subsection{The Io analogy}

A Mercury-like planet is not the only possibility for
\object{CoRoT-7b}. Because the planet is heated not only by the
radiation of the star but also the strong tidal forces, it may resemble
Io.  Barnes et al. (\cite{barnes10}) pointed out, that the tidal
heating may be as strong as the radiative heating by the star, if
the orbit of \object{CoRoT-7b} is slightly eccentric ($e\geq
10^{-5}$). This eccentricity could be driven by
\object{CoRoT-7c}. In this case, \object{CoRoT-7b} may be dominated
by volcanism and rapid resurfacing, and possibly even have a
plasma torus like Io.  These types of planets had
already been suggested before the discovery of \object{CoRoT-7b} by
Briot \& Schneider (\cite{briot08}).  However, even if
\object{CoRoT-7b} had strong volcanic activity, it is not immediately
obvious that it would also have a plasma torus similar to that of Io.  Since the
torus of Io emits lines in the wavelength regime covered by our
spectra, we can also search for them.  Prominent lines originating
from the torus of Io are the Na\,I $D_{1,2}$-lines and the forbidden
emission lines [S\,II] ($\lambda \lambda$ 6716, 6731 \AA), [S\,III]
($\lambda$ 6312 \AA ), and [O\,III] ($\lambda$ 5007 \AA) (e.g. Brown
et al.  \cite{brown75}, Rauer et al. \cite{rauer93}, Thomas
\cite{thomas96}).

\section{Observations}

High-resolution spectroscopy of \object{CoRoT-7b} taken in- and
out-of-transit is ideal for detecting the exosphere or the plasma
torus, because the lines are expected to be very narrow, like those of
Mercury. The broadening of the lines is mainly due to the change in
the radial velocity of the planet during the course of the
observations.

Two sets of observations were carried out in service mode with UVES on
ESO VLT UT-2 (KUEYEN). In the night of 27-28 December 2009,
\object{CoRoT-7b} was observed for 50 minutes during transit and for
50 minutes out-of-transit. The out-of-transit observations were taken
about 3.5 hours after the end of the transit, to make sure that the
tail is no longer in front of the star.  The second data-set was taken
on the 3 January 2010 and 7 January 2010. The in-transit observations
were taken on the 3 January 2010 , the out-of-transit observations on
the 7 January 2010. In each case, the observing time was only 15
minutes. In the following, we indicate the first data set with an $I$,
and the second data set with an $II$.

During our observations, we used the standard 390+580 setting, which
covers the wavelength region from 3270 \AA \, to 6820 \AA.
Unfortunately, orders that are only partly covered by the detector,
cannot be used for this type of analysis.  Thus, the blue channel
effectively covers only the spectral range from 3290 to 4520
\AA. There are two detectors for the red channel, which cover the
wavelength regions from 4780 to 5740, and from 5820 to 6760 \AA .

We used a slit-width of 0.6 arcsec, which provides a resolution of
$\lambda/\Delta \lambda =$ 65000.  Apart from the differences in
exposure-time, the second data-set has the disadvantage that the in-
and out-of-transit observations were taken in different nights.  Given
the difference in exposure-times and signal-to-noise (S/N) ratio, the
data-set $I$ was used to search for the lines from the planet, and
data-set $II$ to confirm them if they were found.  Another interesting
property of data-set $II$ is that its out-of-transit observations were
taken at phase 0.7, compared to 0.2 for the data-set $I$.

A series of IRAF routines were used to remove both the bias
and scattered light, to flat-field the spectra, and then to extract and
wavelength-calibrate them.  The sky background was removed by
extracting the spectrum of the night-sky along the slit and
subtracting it from the spectrum of the star.  For the wavelength
calibration, we used the ThAr spectra taken after the observations and
performed a global fit using 886 and 1523 ThAr lines for the two CCDs
of the red arm and 2518 ThAr lines for the blue arm.  The mid-points
of the observation, and their corresponding phases are given in
Table\,\ref{tab01}. The transit lasts about $1.3\,h$, or $0.0635$ in
phase.

\begin{table*}
\caption{Observing log}
\begin{tabular}{l l l l l l }
\hline
name & date & UT        & exposure  & HJD       & phase     \\
     &      & mid point & time [s] & mid-point & mid-point \\ 
\hline
transit Ia  & 2009.12.28 & 02:43:14 & 970 & 2455193.61851 & 0.9899 \\
transit Ib  & 2009.12.28 & 03:00:06 & 970 & 2455193.63023 & 0.0036 \\
transit Ic  & 2009.12.28 & 03:15:04 & 970 & 2455193.64062 & 0.0158 \\
\hline
off Ia      & 2009.12.28 & 06:45:20 & 970 & 2455193.78664 & 0.1868 \\
off Ib      & 2009.12.28 & 07:02:12 & 970 & 2455193.79835 & 0.2006 \\
off Ic      & 2009.12.28 & 07:19:09 & 970 & 2455193.81012 & 0.2144 \\
\hline
transit II  & 2010.01.03 & 02:29:17 & 970 & 2455199.60885 & 0.0077 \\
off     II  & 2010.01.07 & 02:46.33 & 970 & 2455203.62067 & 0.7076 \\
\hline
\end{tabular}
\label{tab01}
\end{table*}

% transit at: 
% 2455193.627148
% 2455199.602306
% 2455203.016682
% 2455203.870276
% orbital period: 0.853594
% 485 second are in phase 0.00656622 (1/152.06)

\section{Results}

\subsection{Sodium}

\begin{figure}[h]
\includegraphics[width=0.38\textwidth,angle=-90]{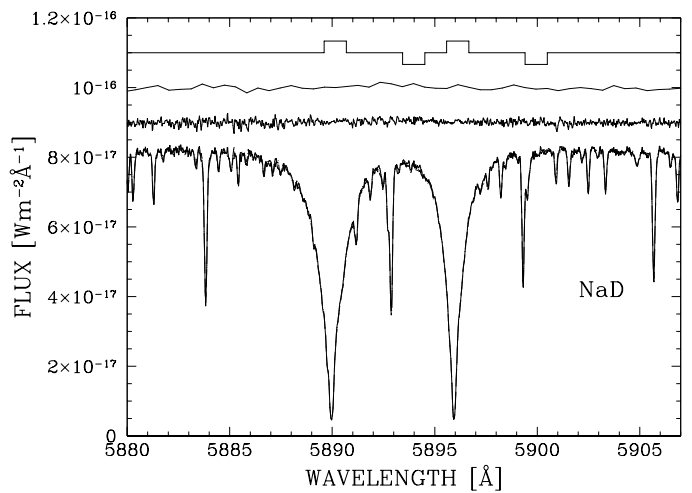}

\caption{Spectra taken in- (full line) and out-of-transit (dashed line) 
  in the range of the Na D-lines. The two lines above the spectrum 
  represent the difference between the two spectra, and the difference after
  resampling to the expected line-width of 55 $km\,s^{-1}$.}
\label{NaD}
\end{figure}

Because the NaD-lines have been detected at Mercury and in the plasma
torus of Io, we select these lines for our study.  Figure\,\ref{NaD}
shows the coadded spectrum of \object{CoRoT-7b} of data-set $I$. The
wavelength is in the rest frame of the star.  The full line is the
spectrum taken in-transit, and the dashed line is the spectrum taken
out-of-transit.  The S/N of the coadded spectra of the first data-set
are about 150 per pixel. Since there are 4 pixel per resolution
element, the resulting S/N per resolution element is 300 for the
data-set $I$, and 180 for data-set $II$.  The quality of the spectra
is so high that it is straightforward to discern even a small difference
between the spectra after resampling the spectra.  The two
NaD-lines are very normal absorption lines as expected for a
G-type star. There is no evidence for any additional emission or
absorption line.

In the next step, we subtracted the out-of-transit spectrum from the
in-transit spectrum (middle line in Fig.\,\ref{NaD}).  The variance in
the difference between the two spectra is $8.5\times 10^{-19}$ 
$W m^{-2} \AA^{-1}$, thus about 100 times smaller than the flux of the
continu\-um in each spectrum, which is $8.2\times 10^{-17}$ $W m^{-2}
\AA^{-1}$.  Because the planet changes its radial velocity by 55
$km\,s^{-1}$ during the exposure, we expect any line emitted by
the planet to have (at least) this width. Thus, we resampled the
spectrum to this resolution, which reduced the noise yet further
(Fig.\,\ref{NaD}.) After resampling, the resolution is reduced to
$\lambda /\Delta \lambda=5500$, and the S/N had increased by a factor
of three. After subtracting the spectra from each other, we then
performed {\em two} different analysis, one to search for narrow lines at full
resolution ($\lambda/\Delta \lambda =$ 65000), and a second to
search for broad features at medium resolution ($\lambda /\Delta
\lambda=5500$).

Do we expect to see an emission or an absorption feature in
Fig.\,\ref{NaD}?  If we define $\alpha$ as the angle
between the line of sight and a tail driven by the radiation of the
star, then $\alpha$ changes from $-7^o$ to $8^o$ during transit
$I$. During the out-of-transit observations for the same night, $\alpha$
changes from $65^o$ to $80^o$. For the data-set $II$, $\alpha$
is $3^o$ for the in-transit observations and $105^o$ for the
out-of-transit observations, respectively.

This means that we observe almost directly along the radiatively
driven sodium-tail during the transit. In the course of the
out-of-transit observations, we see the tail almost from the side. In
both cases, the cross-section of the sodium-tail is not much wider
than the planet (Mura et al. \cite{mura10}). During transit, we expect
to detect the absorption of stellar light, because we are aligned with
the star-planet line (phase angle around $180^o$). Out-of-transit, we
expect to detect emission from the tail of the observations (phase
angle around $90^o$). Thus, if we subtracted the out-of-transit
observations from the in-transit observations, we would also observe
an absorption feature if the planet had a sodium tail best viewed from
the side (out-of-transit).  The line at the top of Fig.\,\ref{NaD}
indicates the expected position and width of either the absorption or
emission feature originating in the planet.  The box pointing
downwards is the position of the sodium lines during the
out-of-transit observations, and the box pointing upwards marks the
expected position of the sodium lines originating in the planet during
transit. A close inspection shows neither an emission nor absorption
feature in Fig.\,\ref{NaD}. Thus, the NaD-lines from \object{CoRoT-7b}
are not detected in the current set of data.

The 3$\sigma$ upper limits of NaD are given in Tables\,\ref{tab02}, and
\,\ref{tab03}.  Assuming that the radiation is emitted spherically
symmetrically, and using the distance of 150 pc, we derive the upper
limits to the total emitted flux for the spectral lines and list them
in the third column of Tables\,\ref{tab02} and \,\ref{tab03}.  Given
the total luminosity of the star of $1.9\times 10^{26}\,W$, we also
derive the upper limits for the lines divided by the luminosity of the
star.  For the NaD-lines, we conclude that we might have detected a
line that either absorbs or emits only $2.2\times 10^{-6}\,L_{*}$ . One 
may argue that if the emission originated in a torus, we would not detect
it after subtracting the out-of-transit spectrum from the in-transit
spectrum, because the emission would originate in both.  However, the
emission along the torus of Io is not constant so would not
cancel out. A line emitted by a ring or a torus that has the same
distance from the star as the planet would have a width of 430
$km\,s^{-1}$.  The corresponding spectral line would thus be very broad and
have the shape of a spectral line like Io (Thomas \cite{thomas96}).
Prior to the subtraction of the two spectra, we inspected the spectra
carefully but did not find any broad emission line. The hypothesis of
a sodium torus is thus excluded.

\begin{table}
\caption{3 $\sigma$-upper limits to the fluxes for the first data set}
\begin{tabular}{l l l l }
\hline
line     & measured       & total & fraction \\
         & $W m^{-2}$      & $W$  & of $L_{*}$\\ 
\hline
Ca II K   & $2.9\times 10^{-18}$ & $7.7\times 10^{20}$ & $4.0\times 10^{-6}$ \\
Ca II H   & $3.2\times 10^{-18}$ & $8.7\times 10^{20}$ & $4.6\times 10^{-6}$ \\
Ca I 4227 & $3.9\times 10^{-18}$ & $1.0\times 10^{21}$ & $5.4\times 10^{-6}$ \\
Na $D_1$  & $1.6\times 10^{-18}$ & $4.2\times 10^{20}$ & $2.2\times 10^{-6}$ \\
Na $D_2$  & $1.6\times 10^{-18}$ & $4.2\times 10^{20}$ & $2.2\times 10^{-6}$ \\
CaO       & $1.0\times 10^{-17}$ & $2.6\times 10^{21}$ & $1.4\times 10^{-5}$ \\
\hline
$[O\,III]$ 5007 & $4.4\times 10^{-18}$ & $1.2\times 10^{21}$ & $6.1\times 10^{-6}$ \\
$[S\,III]$ 6312 & $3.5\times 10^{-18}$ & $9.6\times 10^{20}$ & $5.0\times 10^{-6}$ \\
$[S\,II]$ 6716  & $3.1\times 10^{-18}$ & $8.4\times 10^{20}$ & $4.4\times 10^{-6}$ \\
$[S\,II]$ 6731  & $3.1\times 10^{-18}$ & $8.4\times 10^{20}$ & $4.4\times 10^{-6}$ \\
\hline
\end{tabular}
\label{tab02}
\end{table}

\begin{table}
\caption{3 $\sigma$-upper limits to the fluxes for the second data set}
\begin{tabular}{l l l l }
\hline
line     & measured       & total & fraction \\
         & $W m^{-2}$      & $W$  & of $L_{*}$\\ 
\hline
Ca II K   & $8.5\times 10^{-18}$ & $2.2\times 10^{21}$ & $1.2\times 10^{-5}$ \\
Ca II H   & $1.0\times 10^{-17}$ & $2.8\times 10^{21}$ & $1.4\times 10^{-5}$ \\
Ca I 4227 & $7.7\times 10^{-18}$ & $2.1\times 10^{21}$ & $1.0\times 10^{-5}$ \\
Na $D_1$  & $3.4\times 10^{-18}$ & $9.2\times 10^{20}$ & $4.8\times 10^{-6}$ \\
Na $D_2$  & $3.4\times 10^{-18}$ & $9.2\times 10^{20}$ & $4.8\times 10^{-6}$ \\
CaO       & $3.0\times 10^{-17}$ & $7.8\times 10^{21}$ & $4.1\times 10^{-5}$  \\
\hline
$[O\,III]$ 5007 & $2.1\times 10^{-17}$ & $5.8\times 10^{21}$ & $3.0\times 10^{-5}$ \\
$[S\,III]$ 6312 & $8.8\times 10^{-18}$ & $2.4\times 10^{21}$ & $1.2\times 10^{-5}$ \\
$[S\,II]$ 6716  & $6.1\times 10^{-18}$ & $1.6\times 10^{21}$ & $8.5\times 10^{-6}$ \\
$[S\,II]$ 6731  & $6.1\times 10^{-18}$  & $1.6\times 10^{21}$ & $8.5\times 10^{-6}$ \\
\hline
\end{tabular}
\label{tab03}
\end{table}

\subsection{Neutral Ca}

\begin{figure}[h]
\includegraphics[width=0.38\textwidth,angle=-90]{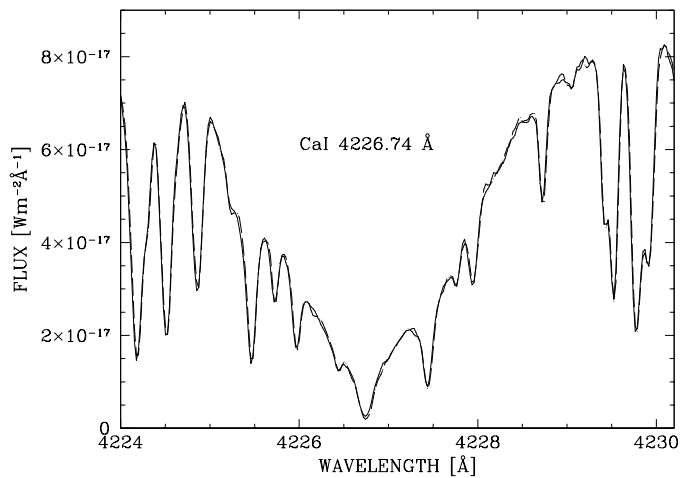}

\caption{Spectra taken in- (full line) and out-of-transit (dashed
  line) in the range of the Ca\,I line at 4226.74 \AA .}
\label{CaI}
\end{figure}

\begin{figure}[h]
\includegraphics[width=0.38\textwidth,angle=-90]{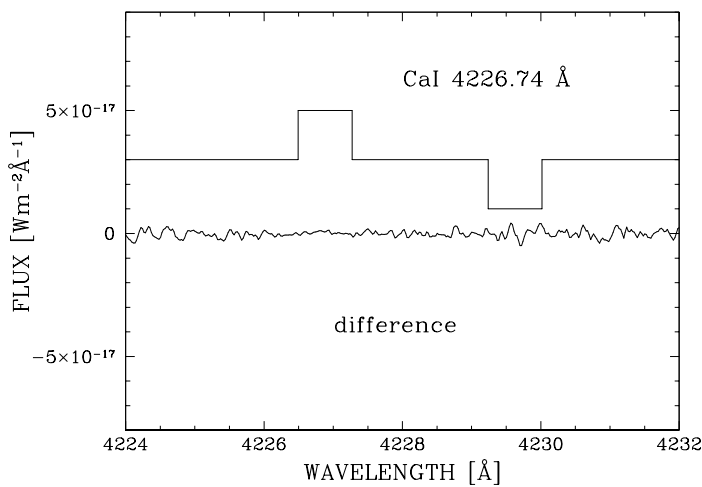}

\caption{Difference between the two spectra shown in
  Fig.\,\ref{CaI}, together with the expected position of the lines
  from CoRoT-7b. Box pointing upwards indicates the expected position 
  of the line during transit. Box pointing downwards indicates the expected 
  position of the line in the out-of-transit observation.}
\label{CaIa}
\end{figure}

Because Bida, Killen \& Morgan (\cite{bida00}) detected the Ca\,I
emission-line at 4226.74 \AA \, we also searched for this line.
Figure\,\ref{CaI} shows the spectra taken in- and out-of-transit, and
Fig.\,\ref{CaIa} their differences.  The line at the top of
Fig.\,\ref{CaIa} shows the expected position and width of the lines
originating from the planet in the same way as shown in
Fig.\,\ref{NaD}.

\subsection{The $Ca^{+}$-ion}

\begin{figure}[h]
\includegraphics[width=0.38\textwidth,angle=-90]{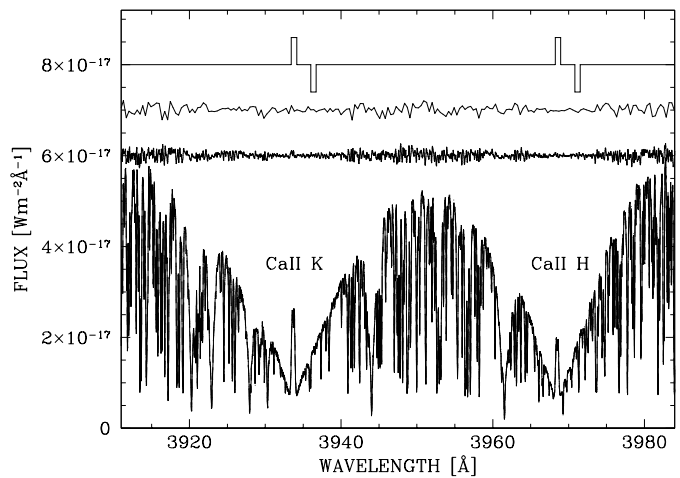}

\caption{Same as Fig.\,\ref{NaD} but for the Ca\,II\,H and K lines.}
\label{CaII}
\end{figure}

Assuming that CoRoT-7b resembles Mercury, a considerable amount of Ca
it is expected in the exosphere, and given the very short photo
ionisation time of Ca it should soon be ionized (Mura et
al. \cite{mura10}). Thus, the Ca II H, K lines are expected to be seen
in the spectra. According to these authors, the $Ca^{+}$-ions would be
guided by the stellar wind and the ion-tail would be at an angle of
about $45^o$ with respect to the star-planet line. Thus, we expect an
absorption feature in the spectrum of the difference
(Fig.\,\ref{CaII}). As before, we resampled the spectrum to a
resolution of 55 $km\,s^{-1}$.  Figure\,\ref{CaII} show the spectra
taken in- and out-of-transit prior to the subtraction.  We again
detect neither an emission nor absorption line originating in the
planet. The upper limits are given in Tables\,\ref{tab02} and
\,\ref{tab03}.

\subsection{The CaO molecule}

\begin{figure}[h]
\includegraphics[width=0.38\textwidth,angle=-90]{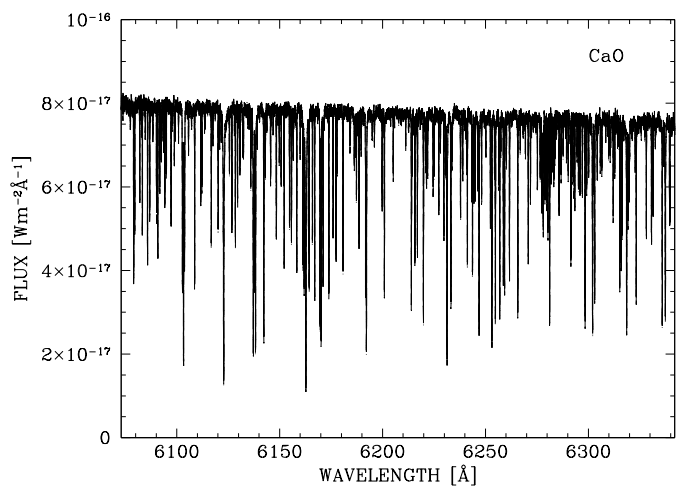}

\caption{Spectra taken in- (full line) and out-of-transit (dashed
  line) in the range of the CaO lines.}
\label{CaO}
\end{figure}

\begin{figure}[h]
\includegraphics[width=0.38\textwidth,angle=-90]{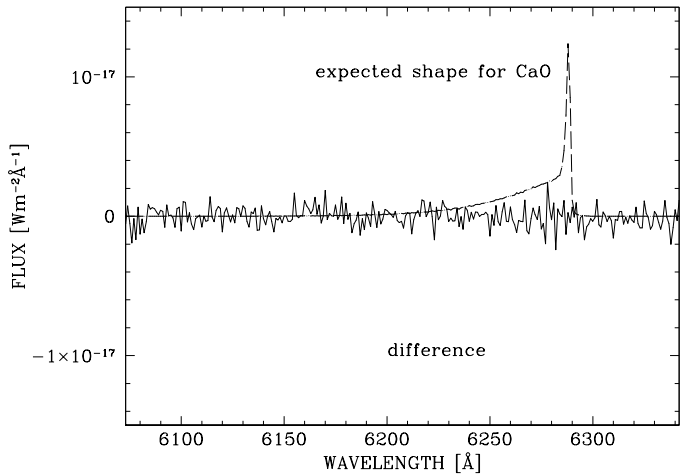}

\caption{Difference of the spectra taken in- and out-of-transit shown
  in Fig.\,\ref{CaO} after resampling the spectrum to a resolution of
  55 $km\,s^{-1}$.  The dashed line demonstrates how the CaO-lines
  would look.}
\label{CaOdiffReb}
\end{figure}

The next series of plots shows the analysis for CaO (Figs.\,\ref{CaO}
and \ref{CaOdiffReb}). Since CaO is a molecule, it has a rich spectrum
of transitions.  Figure\,\ref{CaOdiffReb} shows the calculated CaO
spectrum as dashed line overplotted on the difference spectrum for the
in-transit and out-of-transit observations.  The theoretical CaO
spectrum shown in Fig.\,\ref{CaOdiffReb} was calculated assuming a
temperature of 2500 K and folding the resolution of our UVES
observations. Thus, the individual lines are not resolved but there
should be a very prominent peak, which is not seen. The relatively
high upper limits of $1.0\times 10^{-17}$ and $3.0\times 10^{-17}$ $W
m^{-2}$ is caused by the feature not being very broad, and we assume
that the peak of the feature has to be 3$\sigma$ above the noise
level.

\subsection{The $\rm S^{+}$, $\rm S^{2+}$, and $\rm O^{2+}$-ions}

In addition to the hypothesis that \object{CoRoT-7b} resembles
Mercury, we also tested the hypothesis that more closely resembles Io.
In this case, we would expect to detect NaD-lines but as we have
already seen above, there is no detectable NaD-emission from the
planet.  Other possibilities are $[S\,II]$ ($\lambda \lambda$ 6716,
6731 \AA ), $[S\,III]$ ($\lambda$ 6312 \AA ), and $[O\,III]$
($\lambda$ 5007 \AA ), which we also failed to detect (e.g. Brown et
al.  \cite{brown75}, Rauer et al. \cite{rauer93}, Thomas
\cite{thomas96}).  The upper limits to the fluxes of these lines are
given in Tables\,\ref{tab02} and \,\ref{tab03}.

\subsection{Upper limits for the whole spectrum}

\begin{figure}[h]
\includegraphics[width=0.38\textwidth,angle=-90]{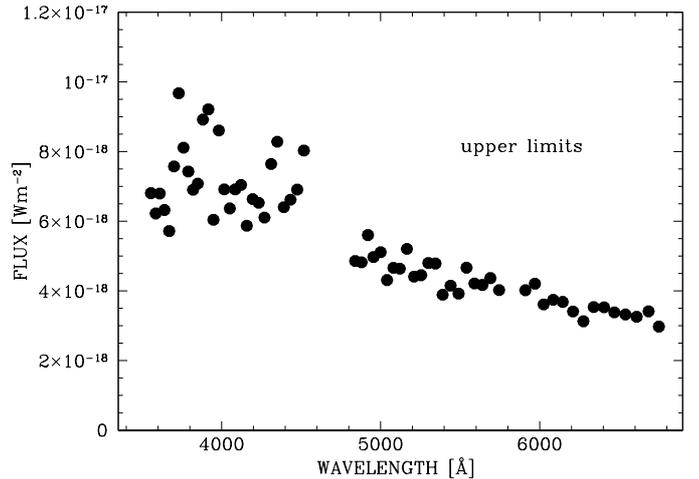}

\caption{3 $\sigma$-upper limits to the fluxes of lines
per spectral order.}
\label{UpperLimits}
\end{figure}

To maximise our chances of detection, we carried out the same kind of
analysis for the whole UVES spectrum. Figure\,\ref{UpperLimits} shows
the upper limits derived for each order.  Because of the
blaze-function of the spectrograph, there are more photons at the
centre of each order than at the edges. This means that the noise
level and the upper limits are higher at the edges of an order than
its centre. We always derived the upper limits for a whole order to
avoid making Fig.\,\ref{UpperLimits} too complicated.  Given the form
of the blaze-function, this effect is greater in the blue than the
red. However, the specific spectral line may be at the centre of one
order and closer to the edges of another one.  Since the overlap of
the orders is larger in the blue than in the red, the effect of the
blaze-function is partly compensated. This explains why the upper
limits for Ca\,II\,H,K are not significantly poorer than those for
NaD.  The result of the analysis is that no emission or absorption
line originating in the planet is found in the whole spectrum.

\section{Discussion and conclusions}

\object{CoRoT-7b} is the first small transiting extrasolar planet with
a density comparable to, or even higher than those of Mercury, Earth,
Venus, and Mars. It is thus reasonable to assume that it is rocky.
The radius of the planet is $1.58\pm0.09\,R_{Earth}$, or
$4.1\pm0.2\,R_{Mercury}$.  Given its small distance from the star, the
surface of the planet is expected to be very hot. L\'eger et
al. (\cite{leger09}) estimated the temperature to be in the range
1800-2600\,K.  Barnes et al. (\cite{barnes10}) pointed out that there
is tidal heating possibly as strong as the heating by the star. These
properties make it reasonable to assume that material from the surface
could be released into space to form an atmosphere as in the case of
Mercury or Io. If this material were detected, we would be able to
perform mineralogy of a rocky, extrasolar planet.  We have searched
for the emission and absorption lines of \object{CoRoT-7b} using
spectra taken in- and out-of-transit. Although our measurements are
very sensitive, we could not detect any spectral lines originating in
the planet.

How well do our upper limits compare to the emission observed for
Mercury? For the $Ca\,I\,4227\,\AA$-line, Bida, Killen \& Morgan
(\cite{bida00}) found a column emission intensity of $384\times 10^6$
$photons\,cm^{-2}\,s^{-1}$, corresponding to $5.1\times
10^{17}\,W$. Our upper limit is $10^{21}\,W$, or about 2000 times
higher than the line-flux detected for Mercury. Since the surface area
of CoRoT-7b is about 17 times larger than that of Mercury, and
\object{CoRoT-7b} receives about 250-370 times more radiation than
Mercury, our upper limit is even lower than the Ca-emission observed
at Mercury if we scale accordingly. However, since the photoionisation
of Ca atoms is very fast for CoRoT-7b, we expect to observe
$Ca^+$-emission, which is not the case.

Given the extreme conditions on this planet and the sensitivity of our
measurements, it is surprising that we have been unable to detect the
exosphere.  On the other hand, since we do not have a planet like this
in our solar system, we have had to make a number of assumptions. We
assumed for example that one side of the planet is always facing the
star. If this were not the case, the conditions would be less extreme
than we think.  Our knowledge of the stellar wind for this star is
also very limited. Although we are certain that this star has a wind,
and we are also certain that the plasma flow will dictate the flow of
ions originating in the planet, we know neither the velocity nor
density of the stellar wind. Thus, we do not know the volume shape
of the interacting region between the stellar plasma and the ions from
the planet.

We have also assumed that the surface of \object{CoRoT-7b} is similar
to that of Mercury.  In this case, lines from Ca\,II and Na would be
expected. But how well is this assumption justified? At the moment,
all we know is that the measured density corresponds to a planet with
a composition similar to earth.  If the surface composition were very
different, it would not be surprising if we did not detect these
lines. This is a possibility, since Schaefer \& Fegley
(\cite{schaefer09}) suggested that "super-Earths" do not have large
quantities of Ca and Na. In this case, the planet should form a
Mg-ion-tail, which is not detectable in the optical regime.  Schaefer
\& Fegley (\cite{schaefer09}) considered a surface composition for
super-Earth's that evolved in a different way to Mercury's. Their
model calculations indicate that volatile elements such as H, C, N, S,
and Cl have been lost by the planet, but that silicate atmospheres
composed primarily of Na, $O_2$, O, Mg, and SiO gas may remain.  The
major atmospheric compounds are most likely to be O, Mg, and SiO. The
atmospheric composition may be altered by fractional vaporization,
cloud condensation, photoionization, and a reaction with any residual
volatile elements remaining in the atmosphere.

We also exclude the presence of a sodium torus like Io.  The torus of
Io around Jupiter may survive because there is the corotating plasma of
Jupiter's plasma sphere. For \object{CoRoT-7b} the ionized sodium
would be quickly carried away by the stellar wind, so that it is
unlikely that a torus can form. It is thus not surprising that we did
not find a sodium torus.

In any case, for the first time we have been able to place constraints on the
properties of an exosphere of a rocky extrasolar planet outside our
solar system. This is particularly interesting, because
\object{CoRoT-7b} does not resemble any of the rocky planets in our
solar-system. Our observations also raise the question of how well it is
justified to assume that this planet is a scaled-up version of
objects in our solar system.

\begin{acknowledgements}
  We are grateful to the User Support Group of ESO/Paranal. Part of
  this work was supported by a grant from the Deutschen Zentrums f\"ur
  Luft- und Raumfahrt (DLR) (50OW0204).  H. Lammer, M. Fridlund,
  J. Schneider, A. Mura, and H. Rauer, also acknowledge the
  International Space Science Institute (ISSI; Bern, Switzerland) and
  the ISSI team "Evolution of Exoplanet Atmospheres and their
  Characterization". The authors also acknowledge support by the
  Europlanet FP7 project and fruitful discussions within the Na2
  working groups WG4 and WG5.
\end{acknowledgements}

\end{document}